# Chemical trends in the high thermoelectric performance of the pyrite-type dichalcogenides: $ZnS_2$, $CdS_2$ and $CdSe_2$

Tiantian Jia[1,2,4], Jesús Carrete[3], Georg K.H. Madsen[3], Yongsheng Zhang[1,5,*] and Suhuai Wei[2,*]

1) Key Laboratory of Materials Physics, Institute of Solid State Physics, HFIPS, Chinese Academy of Sciences, 230031 Hefei, China

2) Beijing Computational Science Research Center, 100193 Beijing, China

3) Institute of Materials Chemistry, TU Wien, A-1060 Vienna, Austria

4) University of Science and Technology of China, 230026 Hefei, China

5) Advanced Research Institute of Multidisciplinary Sciences, Qufu Normal University, 273165 Qufu, China

Corresponding author:

yshzhang@theory.issp.ac.cn (Yongsheng Zhang)

suhuaiwei@csrc.ac.cn (Suhuai Wei)

## ABSTRACT

The thermoelectric properties of the three pyrite-type IIB-VIA$_2$ dichalcogenides ($ZnS_2$, $CdS_2$ and $CdSe_2$) are systematically investigated and compared with those of the prototype $ZnSe_2$ in order to optimize their thermoelectric properties. Using the phonon Boltzmann transport equation, we find that they all have ultralow lattice thermal conductivities. By analyzing their vibrational properties, these are attributed to soft phonon modes derived from the loosely bound rattling-like metal atoms and to strong anharmonicities caused by the vibrations of all atoms perpendicular to the strongly bound nonmetallic dimers. Additionally, by correlating those properties along the series, we elucidate a number of chemical trends. We find that heavier atom masses, larger atomic displacement parameters and longer bond lengths between metal and nonmetal atoms can be beneficial to the looser rattling of the metal atoms and therefore lead to softer phonon modes, and that stronger nonmetallic dimer bonds can boost the anharmonicities, both leading to lower thermal conductivities. Furthermore, we find that all three compounds have complex energy isosurfaces at valence and conduction band edges that simultaneously allow for large density-of-states effective masses and small conductivity effective masses for both p-type and n-type carriers. Consequently, the calculated thermoelectric figures of merit ($ZT$), can reach large values both for *p*-type and *n*-type doping. Our study illustrates the effects of rattling-like metal atoms and localized nonmetallic dimers on the thermal transport properties and the importance of different carrier effective masses to electrical transport properties in these pyrite-type dichalcogenides, which can be used to predict and optimize the thermoelectric properties of other thermoelectric compounds in the future.

# INTRODUCTION

The thermoelectric (TE) conversion efficiency of a device depends on the dimensionless figure of merit, $ZT = S^2\sigma T/\kappa$, where $S$ is the Seebeck coefficient, σ is the electrical conductivity, $T$ is the absolute temperature, and $\kappa$ is the thermal conductivity including charge carrier ($\kappa_e$) and lattice ($\kappa_l$) contributions. Promising TE materials must have high thermopower factors ($PF = S^2\sigma$) and low thermal conductivities. In semiconductors, the thermal conductivity is mainly determined by phonon properties that control the lattice contribution. However, since the state-of-the-art TE materials still have relatively low $ZT$ values or contain expensive elements, the commercial applications of TE devices are limited. Two major strategies are currently used to increase the $ZT$ values of TE materials: using band structure engineering to improve the $PF$[1–3] and/or increasing phonon scatterings to decrease $\kappa_l$.[4–6] Additionally, owing to the development of the associated computational methodologies, high-throughput methods have been used to seek high-performance TE materials as well.[7–10]

Recently, the pyrite-type compounds $FeX_2$ (X=S, Se and Te)[11–13], $PtSb_2$[14] and $MnTe_2$[15], have shown promising thermoelectric performance based on either first-principles studies or experimental measurements. Moreover, through our previous high-throughput calculations[16], we also predicted that a distinct pyrite-type IIB-VIA$_2$ dichalcogenide ($ZnSe_2$) has an excellent thermoelectric performance, due to the complex electronic energy isosurfaces and the existence of the rattling-like Zn atoms around the localized Se-Se dimers that improve the electronic and lattice contributions respectively.[17] However, the thermoelectric properties of many other pyrite-type dichalcogenides with the same prototype still remain to be investigated. Therefore, we set out to explore the thermoelectric performances of three other pyrite-type IIB-VIA$_2$ dichalcogenides: $ZnS_2$, $CdS_2$ and $CdSe_2$, which were successfully synthesized at high pressures (6.5-8.9 GPa) in 1968,[18] and to compare their properties with $ZnSe_2$ to identify the chemical trends.

The crystal structure of the pyrite-type dichalcogenides ($MX_2$, M=transition metal, X=S and Se) is cubic and belongs to the $Pa\bar{3}$ space group (Fig. 1). In this work, we investigate the thermal and electrical transport properties of three pyrite-type IIB-VIA$_2$ dichalcogenides ($ZnS_2$, $CdS_2$ and $CdSe_2$), and evaluate their thermoelectric performances. By calculating and analyzing their phonon properties, we find that similar to the previously studied $ZnSe_2$,[17] all three pyrite-type dichalcogenides have easily identifiable localized high-frequency optical phonons contributed by their strongly covalently bound nonmetallic dimers and soft phonon modes contributed by their rattling metal atoms. Furthermore, they all have strong anharmonicities and low thermal conductivities [below 2 W/(m·K) at 300 K]. Additionally, we find some chemical trends in that (i) Heavier atom masses, larger atomic displacement parameters and longer bond lengths between metal and nonmetal atoms can result in the softer phonon frequencies; (ii) The vibrations of all atoms perpendicular to the nonmetallic dimers can contribute to their strong anharmonicities, and the stronger nonmetallic dimer bonding is, the more obvious the anharmonicity. Furthermore, we find that all three compounds show promising electrical transport properties for both p-type and n-type doping, owing to their large density-of-states effective masses and small conductivity effective masses, which can be explained in terms of by the complex non-spherical isoenergy Fermi surfaces in both the valence and conduction bands. Finally, the low lattice thermal conductivity and promising electrical transport properties contribute to their excellent thermoelectric performances. Our work can help and guide experimental researchers to find new high-performance thermoelectric materials.

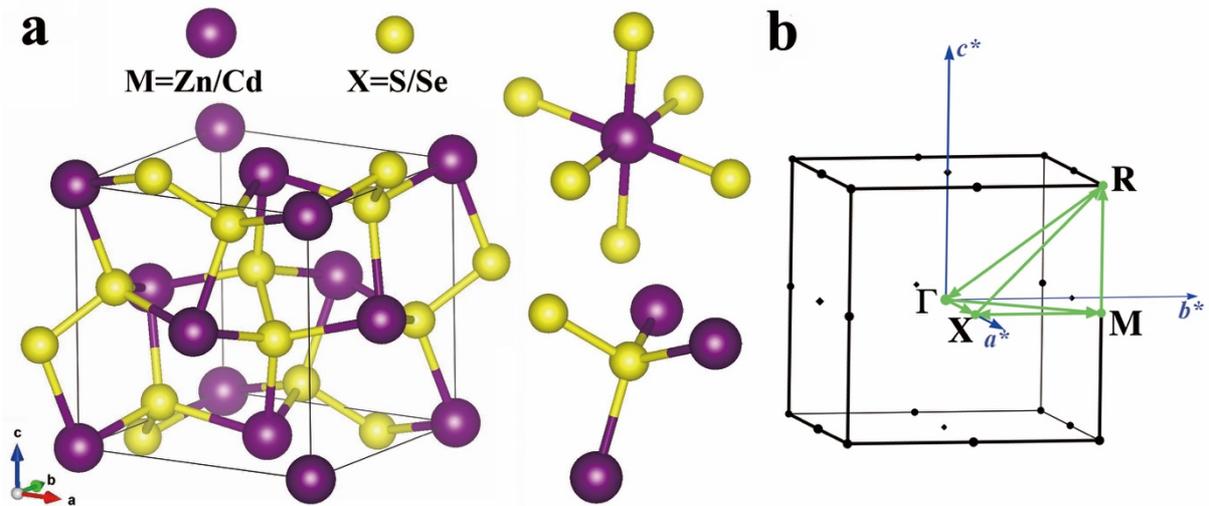

Figure 1. (a) Crystal structures and bonds between atoms in the pyrite-type IIB-VIA$_2$ dichalcogenides [$Pa\bar{3}$-MX$_2$, M=transition metal (Zn and Cd), X=S and Se]. The Wyckoff position are 4a (0,0,0) for M and 8c ($u,u,u$) for X, where $u$ is the Wyckoff pyrite parameter. In this structure, each M atom has six X neighbors, whereas each X atom binds with three M and one X atoms, and the neighboring X atoms form the X-X dimers. (b) The corresponding first Brillouin zone and its high-symmetry points.

## COMPUTATIONAL METHODOLOGY

The electronic properties are calculated by means of density-functional theory (DFT) as implemented in the Vienna Ab initio Simulation Package (VASP)[19] with the projector augmented-wave method.[20] Their electronic exchange and correlation energies are accounted for through the generalized gradient approximation of Perdew, Burke, and Ernzerhof (GGA-PBE).[21] The energy cutoffs for the plane-wave expansion are 420 eV. To calculate the phonon vibrational properties, the crystal structures are relaxed with convergence parameters: a $8 \times 8 \times 8$ k-point mesh, the convergence of the total energies to less than $10^{-8}$ eV and the force components of each atom below $10^{-4}$ eV/Å. In addition, $2 \times 2 \times 2$ supercells are used for the second- and third-order interatomic force constants (IFCs) calculations by resolving the irreducible set of atomic displacements, as implemented in the PHONOPY software package[22] and the thirdorder.py script.[23] Additionally, the Born effective charges and dielectric tensors are also extracted from the DFT calculations in order to apply a long-range correction to the phonon dispersions.

The lattice thermal conductivities ($\kappa_l$) as functions of temperature ($T$) are calculated by solving full linearized phonon Boltzmann transport equation (BTE) using the almaBTE code[24] and applying the formula

$$\kappa_l^{\alpha\beta} = \sum_s \sum_q \hbar \omega_{sq} \frac{\partial f_{0,ph}}{\partial T} v_\alpha^p(\omega_{sq}) v_\beta^p(\omega_{sq}) \tau^p(\omega_{sq}) \qquad (1)$$

where $\alpha$ and $\beta$ ($\alpha, \beta = x, y, z$) are the tensor indices, $s$ is the index of phonon branches, $q$ is the wave vector, $\hbar$ is the reduced Planck constant, $\omega$ is the phonon frequency, $f_{0,ph}$ is the Bose-Einstein distribution function, $v^p(\omega)$ is the phonon velocity, and $\tau^p(\omega)$ is the phonon relaxation time. Interatomic interactions up to the 5$^{th}$ coordination shell are considered, and a $7 \times 7 \times 7$ Γ-centered grid is used to solve the phonon BTE. The electrical transport properties ($\sigma/\tau$, $S$ and $\kappa_e$) as functions of temperature and chemical density are calculated using the BoltzTraP2 code.[25] Additionally, since the carrier relaxation time ($\tau$) at room temperature is typically $10^{-14}$ to $10^{-15}$ s and almost energy-independent for most semiconductors,[26] the constant relaxation time approximation (CRTA) has been a staple of many previous electrical transport calculations related to thermoelectricity.[27–29] Therefore, in this work, the electrical transport properties of the three pyrite-type dichalcogenides are evaluated by the simple CRTA.

## RESULTS AND DISCUSSION

### A. Lattice parameters and phonon properties

For the three pyrite-type IIB-VIA$_2$ dichalcogenides: ZnS$_2$, CdS$_2$ and CdSe$_2$, the calculated lattice constants ($a$) are 5.9965, 6.3990 and 6.7403 Å, which are in reasonable agreement with the experimental measurements (5.9542, 6.3032 and 6.615 Å),[18] and the calculated pyrite parameters ($u$) are 0.400, 0.406 and 0.397, consistent with the previous theoretical values.[30]

To evaluate their structural stabilities under ambient pressure, we calculate their phonon dispersions along different high symmetry lines (refer to the Brillouin zone in Fig. 1b), as shown in Fig. 2. Three acoustic phonon branches [two transverse (TA, TA') and one longitudinal (LA)] are highlighted. From Fig. 2, we can see that all phonon frequencies of the three dichalcogenides are real (depicted as positive), which indicates that all of them are mechanically stable at zero pressure.

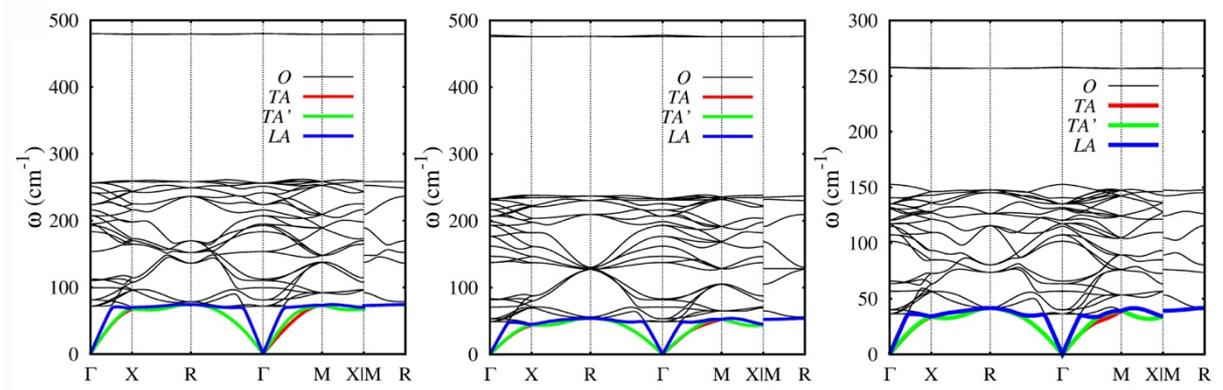

Figure 2. Calculated phonon dispersions of $ZnS_2$, $CdS_2$ and $CdSe_2$ (from left to right), respectively. The red, green and blue lines highlight two transverse (TA, TA') and one longitudinal (LA) acoustic phonon modes, respectively. The black lines represent optical (O) phonon modes.

In most semiconductors, the thermal conductivities are mainly contributed by phonons. By comparing the shapes and values of the calculated phonon dispersions in the three dichalcogenides (Fig. 2), we find that they all have qualitatively similar phonon dispersion curves along different high symmetry lines, and that their phonon frequencies ($\omega$) generally follow $\omega(ZnS_2) > \omega(CdS_2) > \omega(CdSe_2)$, with maximum values for acoustic phonons of 74.1, 53.9 and 42.3 cm$^{-1}$, respectively. Consequently, their speeds of sound also follow the order $v(ZnS_2) > v(CdS_2) > v(CdSe_2)$. These results are consistent with the inverse proportionality between the speed of sound and the square root of the density,[31] which in this case is dominated by the atomic mass: Cd is heavier than Zn and Se is heavier than S. Furthermore, we also find that they all have isolated, localized high-frequency phonons and obvious bandgaps in the optical phonon regions. The ranges of those bandgaps are from 262 cm$^{-1}$ to 478 cm$^{-1}$ for $ZnS_2$, from 239 cm$^{-1}$ to 475 cm$^{-1}$ for $CdS_2$, and from 153 cm$^{-1}$ to 257 cm$^{-1}$ for $CdSe_2$, respectively, see Fig. 2.

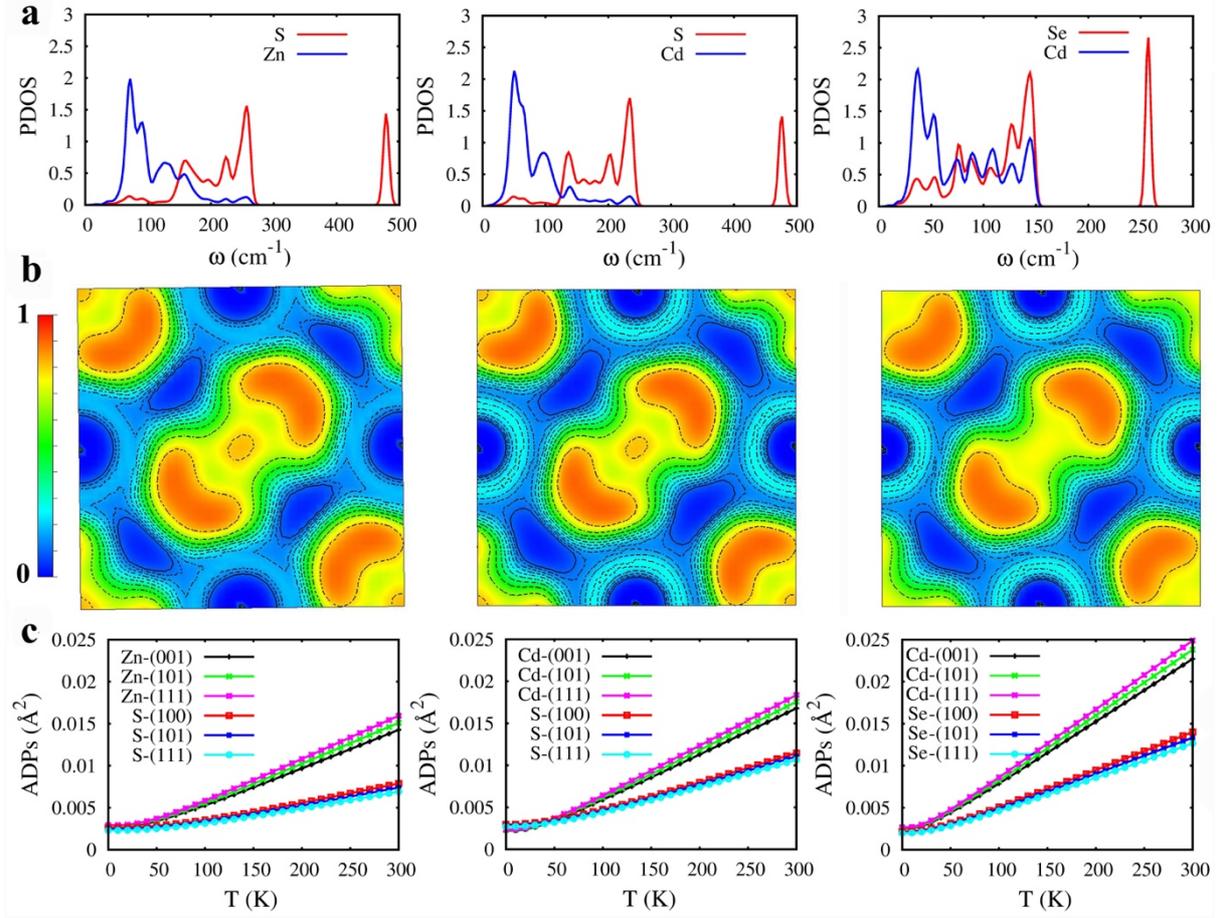

Figure 3. Calculated partial phonon densities of states (PDOS) (a), projected 2D electron localization function (ELF) in (001) plane (b) and atomic displacement parameters (ADPs) along different directions of different atoms (c) in $ZnS_2$, $CdS_2$ and $CdSe_2$ (from left to right), respectively.

By calculating the phonon densities of states (PDOS) of the three dichalcogenides (Fig. 3a), we find that their isolated localized high-frequency optical phonons are only contributed by the nonmetal (X) atoms and the low-frequency phonons are contributed by the metal (M) and nonmetal (X) atoms together. To analyze the reasons, we calculate their projected 2D electron localization function (ELF)[32] in (001) plane as shown in Fig. 3b. The ELF is a measure of the extent of electronic spatial localization at a given position. Higher ELF value indicates more electronic localization (as covalent bonds, core shells, or lone pairs).[32] From Fig. 3b, we find that the calculated ELF values around two adjacent nonmetal (X) atoms are very large, which indicates that the three dichalcogenides all have the strong covalently bound nonmetallic (X-X) dimers. By analyzing the atomic vibrations, we find that their isolated localized high-frequency optical phonons mostly correspond to stretches of the strong nonmetallic dimers, similar to the case of $ZnSe_2$.[17] This means that the strong nonmetallic dimers can result in the isolated localized high-frequency optical phonons in these dichalcogenides.

Low-frequency phonons in the three dichalcogenides [such as those with phonon frequencies ($\omega$) lower than 262, 239 and 153 cm$^{-1}$ in ZnS$_2$, CdS$_2$ and CdSe$_2$, respectively, in Fig. 3a], can be further separated into two regions: below 147, 123 and 73 cm$^{-1}$ in ZnS$_2$, CdS$_2$ and CdSe$_2$, respectively, most of the contribution comes from M; at higher frequencies, the vibrations of X atoms dominate. By calculating the atomic displacement parameters (ADPs) of different atoms, we find that the calculated ADPs of M atoms are much larger than that of X atoms in all three dichalcogenides, and the calculated ADPs of different M atoms follow ADPs(ZnS$_2$_Zn) < ADPs(CdS$_2$_Cd) < ADPs(CdSe$_2$_Cd), as shown in Fig. 3c. This means that the M atoms are loosely bound with X atoms and behave as rattling atoms[33] in all three dichalcogenides and that the rattling behaviors in CdSe$_2$ are more obvious than that in CdS$_2$ and ZnS$_2$, which can be used to respectively explain the lower $\omega$ of M atoms than that of X atoms and the lower $\omega$ in CdSe$_2$ than that in CdS$_2$ and ZnS$_2$ in a way. Additionally, by calculating the bond lengths ($L_{M-X}$) between the rattling metal atom (M) and nonmetal atom (X) (as $L_{Zn-S} = 2.544$ Å, $L_{Cd-S} = 2.733$ Å and $L_{Cd-Se} = 2.851$ Å, which follow $L_{Zn-S} < L_{Cd-S} < L_{Cd-Se}$), and comparing with the calculated phonon properties ($\omega$ and $v$) in the three compounds, we can conclude that the $L_{M-X}$ have an opposite tendency to the calculated $\omega$ and $v$. This means that the longer $L_{M-X}$ can contribute to the softer phonon properties, and another main reason for the lower $\omega$ and smaller $v$ of CdSe$_2$ may be the longer $L_{M-X}$ of CdSe$_2$ than CdS$_2$ and ZnS$_2$.

**B.    Anharmonicities**

Besides the phonon frequency and sound velocity, many previous studies have shown that the anharmonicity (the Grüneisen parameter ($\gamma = -\frac{V}{\omega}\frac{\partial \omega}{\partial V}$) is an important descriptor of $\kappa_l$.[34,35] Therefore, we calculate and compare the values of $\gamma$ at different phonon modes in the three dichalcogenides (as shown in Fig. 4). From the figure, we can find their calculated $\gamma$ follows $\gamma(ZnS_2) > \gamma(CdS_2) > \gamma(CdSe_2)$, and the maximum Grüneisen parameters ($\gamma_m$) in acoustic phonons are 4.60, 3.18 and 2.72 for ZnS$_2$, CdS$_2$ and CdSe$_2$, respectively, which are comparable with other excellent TE materials with strong anharmonicity (such as, AgSbTe$_2$ and SnSe, their Grüneisen parameters being 2.05[36] and 4.1[37], respectively). This means that, like ZnSe$_2$ (whose calculated $\gamma_m$ is 4.6[17]), the three dichalcogenides also show strong anharmonicity.

To explore the origins of the strong anharmonicity in these dichalcogenides, we analyze and compare the calculated $\gamma$ at different phonon modes in the same compound (in Fig. 4). From the figure, we find that although the calculated $\omega$ of two transverse acoustic phonon branches (TA and TA') along the ΓM line in the same compound are very close, the calculated $\gamma$ of those TA and TA' branches are obviously different (the calculated $\gamma$ of the TA branch is far higher than those of the TA' branch). By analyzing the vibration modes of atoms in those TA and TA' branches along the ΓM line (as shown in Fig. S1), we find that the vibrations of all atoms in the TA branch are primarily perpendicular to the nonmetallic dimers but the angles between the vibrations of all atoms in the TA' branch and the direction of the nonmetallic dimers are close to 54.7°. Therefore, combining that with the analysis of $ZnSe_2$, we can infer that the vibrations of all atoms perpendicular to the bonding direction of the strong covalently bound nonmetallic dimers can result in a large change of the pyrite parameters $(u)$[17] and the strong anharmonicity at the acoustic phonons, and the existence of the strong covalently bound nonmetallic dimer has an important influence on the strong anharmonicity in these pyrite-type dichalcogenides. Furthermore, by comparing the bond lengths between two adjacent S atoms in S-S dimers ($L_{S-S}$) in $ZnS_2$ and $CdS_2$, we find that their $L_{S-S}$ are 2.079 and 2.086 Å, respectively. The $L_{S-S}$ in $ZnS_2$ is shorter than $CdS_2$, which means that the bond strengths of the nonmetallic dimers in $ZnS_2$ are stronger than $CdS_2$. Comparing the calculated $L_{S-S}$ with the calculated $\gamma$, we can infer that the stronger nonmetallic dimers are beneficial to the stronger anharmonicity in these pyrite-type dichalcogenides.

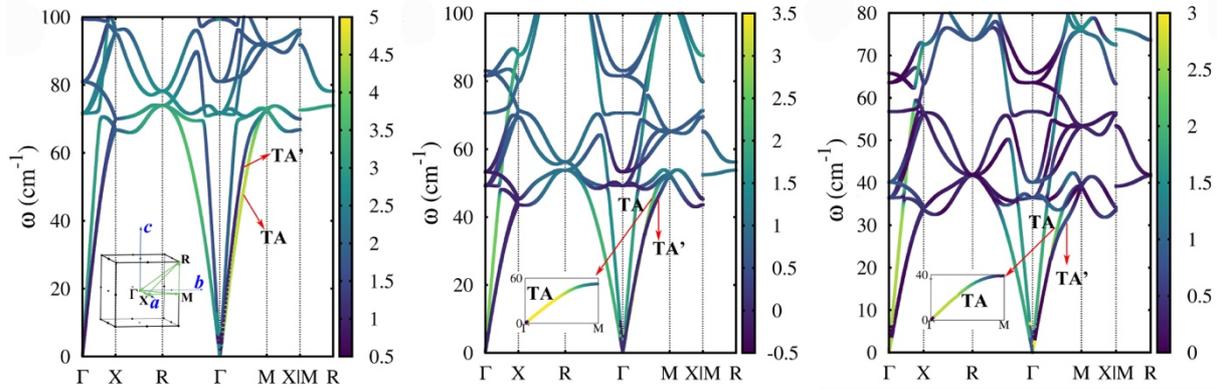

Figure 4. Calculated anharmonicity parameters ($\gamma$) at different phonon modes (the color denotes the values of $\gamma$) for $ZnS_2$, $CdS_2$ and $CdSe_2$ (from left to right), respectively.

## C. Lattice thermal conductivities

From the analyses above, we can see that the three dichalcogenides ($ZnS_2$, $CdS_2$ and $CdSe_2$) all have soft phonon modes and strong anharmonicity, which can contribute to their low lattice thermal conductivity ($\kappa_l$). To verify it, we calculate the $\kappa_l$ of $ZnS_2$, $CdS_2$ and $CdSe_2$ at different temperatures. The results, presented in Fig. 5, reveal that the calculated $\kappa_l$ at 300

K are 1.55, 1.47 and 0.75 W/(m·K) for $ZnS_2$, $CdS_2$ and $CdSe_2$, respectively. By comparing the calculated $\kappa_l$ with the calculated minimum lattice thermal conductivity ($\kappa_{min}$) based on the Ioffe-Regel criterion[38] [the calculated $\kappa_{min}$ at 300 K are 0.13, 0.10 and 0.07 W/(m·K) for $ZnS_2$, $CdS_2$ and $CdSe_2$, respectively], we find that the calculated $\kappa_l$ are much higher the calculated $\kappa_{min}$, which supports to our choice of the phonon BTE method. Additionally, the calculated $\kappa_l$ are lower than the thermal conductivity of PbTe with good thermoelectric performance (2.4 W/mK)[39], which means that all the three dichalcogenides have low lattice thermal conductivities, and consistent with our analyses of their phonon properties. Furthermore, since the heavy atom masses, large atomic displacement parameters and long bond lengths between metal and nonmetal atoms can contribute to the soft phonon properties, the speeds of sound in the three dichalcogenides follow $v(ZnS_2) > v(CdS_2) > v(CdSe_2)$. Since the strong nonmetallic dimers can be beneficial to the strong anharmonicities, their anharmonicities follow $\gamma_m(ZnS_2) > \gamma_m(CdS_2) > \gamma_m(CdSe_2)$. As a result, according to the approximate relationship between $\kappa_l$ and $(v, \gamma)$ as $\kappa_l \propto \frac{v^3}{\gamma^2}$,[40] their calculated lattice thermal conductivities have the trend $\kappa_l(ZnS_2) \sim \kappa_l(CdS_2) > \kappa_l(CdSe_2)$.

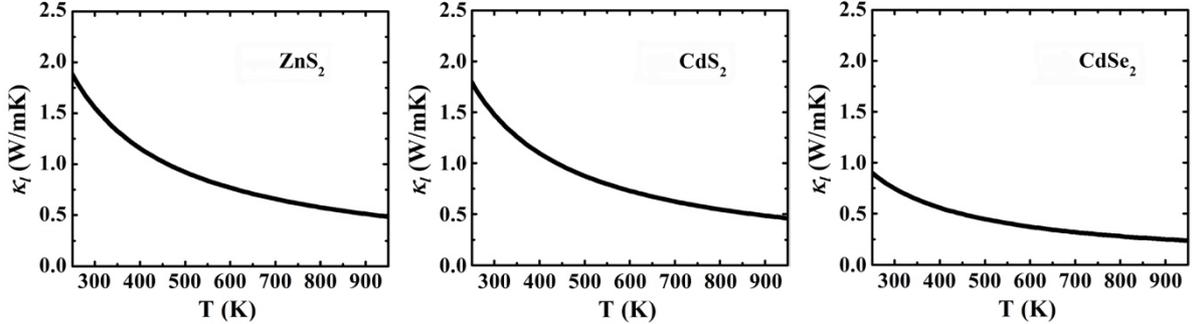

Figure 5. Calculated lattice thermal conductivity ($\kappa_l$) at different temperatures for $ZnS_2$, $CdS_2$ and $CdSe_2$ (from left to right), respectively.

### D. Electronic band structures

To explore the thermoelectric performance of the three dichalcogenides ($MX_2$), we also calculate their electronic band structures and evaluate their electrical transport properties. Fig. 6a shows the calculated PBE electronic band structures of $ZnS_2$, $CdS_2$ and $CdSe_2$ from left to right, respectively. From Fig. 6a, we find that the three iso-electronic compounds have very similar band structure shapes near their Fermi levels. Their valence band maxima (VBM) are also all located at the Γ point with two-fold band degeneracy (as shown in Fig. 6a as B1, B2) and their conduction band minima (CBM) are all located at the P point along the R-Γ line (as shown in Fig. 6a) with a resulting large degeneracy (8) due to symmetry. Their calculated indirect band gaps are 1.41, 1.15 and 0.59 eV, respectively. By calculating their partial charge densities at VBM and CBM (as shown in Fig. S2), we find that their electronic states at the VBM are mainly contributed by the p-electrons in nonmetal atoms ($X_p$) and the electronic states at CBM are mainly contributed by the antibonding states of s-electrons in metal atoms ($M_s$) and s-electrons in nonmetal atoms ($X_s$). Since the calculated atomic energy levels[41,42] of the outermost valence p-electrons are $E(S_{3p}) = -7.112$ and $E(Se_{4p}) = -6.660$ eV, i.e., $E(S_{3p}) < E(Se_{4p})$, the energy levels at the VBM in $MS_2$ will be lower than those in $MSe_2$, which can result in the bigger band gap of $MS_2$ than $MSe_2$. Moreover, since the Zn-S bond lengths in $ZnS_2$ ($L_{Zn-S} = 2.544$ Å) are shorter than the Cd-S bond lengths in $CdS_2$ ($L_{Cd-S} = 2.733$ Å), the antibonding strength of $Zn_{4s}$-$S_{3s}$ in $ZnS_2$ is stronger than $Cd_{5s}$-$S_{3s}$ in $CdS_2$, which can cause the energy levels at CBM in $ZnS_2$ to be higher than in $CdS_2$ and the band gap in $ZnS_2$ to be bigger than in $CdS_2$.

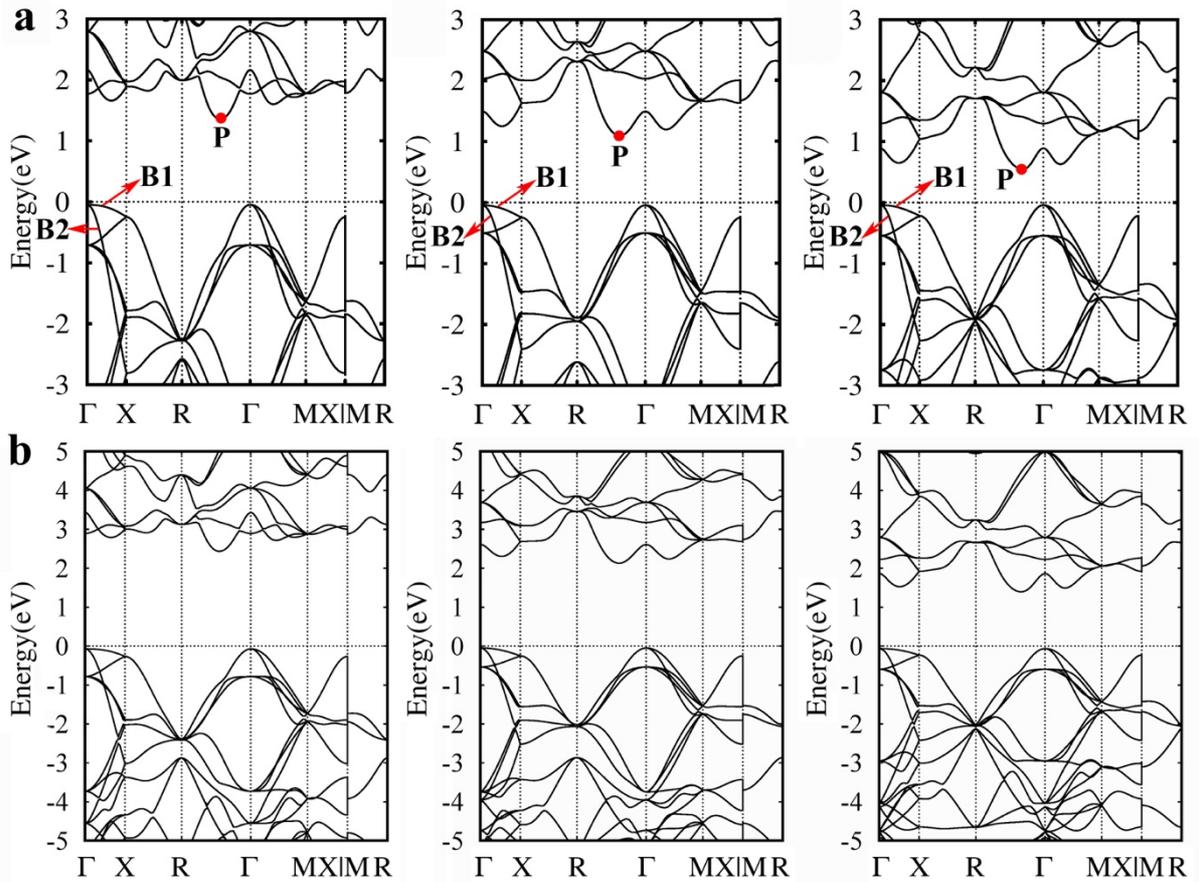

Figure 6. The PBE-calculated (a) and the HSE06-calculated (b) electronic band structures of $ZnS_2$, $CdS_2$ and $CdSe_2$ (from left to right), respectively.

In addition, we compare the band structure shapes of the heavy (B1) and light (B2) bands along the Γ–X direction in the same compound, and we find that the calculated carrier effective masses ($m^*_{\alpha\beta} = \hbar^2 \left[\frac{\partial^2 E(k)}{\partial k_\alpha \partial k_\beta}\right]^{-1}$) along the Γ-X direction are $7.38 m_0$, $4.51 m_0$ and $5.47 m_0$ for B1 bands, and $0.24 m_0$, $0.27 m_0$ and $0.22 m_0$ for B2 bands in ZnS$_2$, CdS$_2$ and CdSe$_2$, respectively (as listed in Table 1, where $m_0$ is the electron rest mass), which means that the three dichalcogenides can provide both light and heavy p-type carriers simultaneously. However, by comparing the calculated $m^*_{\alpha\beta}$ at CBM in the same compound along different directions (Table 1), we find that all the calculated $m^*_{\alpha\beta}$ at CBM are similar and smaller than $m_0$, which means that the three dichalcogenides can provide light n-type carriers and large symmetry degeneracy (8) simultaneously.

Table 1. The carrier effective mass ($m^*_{\alpha\beta}$) of different energy valleys along different directions ($\vec{D}$) for ZnS$_2$, CdS$_2$ and CdSe$_2$, respectively.

| P | $\vec{D}$ | $m^*_{\alpha\beta}$ ($m_0$) | | | | | | | | |
|---|---|---|---|---|---|---|---|---|---|---|
| | | ZnS$_2$ | | | CdS$_2$ | | | CdSe$_2$ | | |
| | | VBM | | CBM | VBM | | CBM | VBM | | CBM |
| | | B1 | B2 | | B1 | B2 | | B1 | B2 | |
| Γ | $\vec{\Gamma X}$ | 7.38 | 0.24 | | 4.51 | 0.27 | | 5.47 | 0.22 | |
| | $\vec{\Gamma M}$ | 0.86 | 0.31 | | 0.91 | 0.36 | | 0.79 | 0.29 | |
| | $\vec{\Gamma R}$ | 0.46 | 0.46 | | 0.51 | 0.51 | | 0.43 | 0.43 | |
| P | $\vec{P_x}$ | | | 0.50 | | | 0.53 | | | 0.41 |
| | $\vec{P_{xy}}$ | | | 0.51 | | | 0.54 | | | 0.43 |
| | $\vec{P\Gamma}$ | | | 0.51 | | | 0.55 | | | 0.45 |

**E.  Electrical transport properties**

Since the PBE is well-known for underestimating the band gaps of semiconductors and insulators, we further calculate the electronic band structures and band gaps of the three dichalcogenides using the Heyd–Scuseria–Ernzerhof (HSE06)[43,44] method (as shown in Fig. 6b). Comparing Fig. 6b and Fig. 6a, we find that all the HSE06-calculated band features near the Fermi level are similar with the PBE results, and the HSE06-calculated band gaps are bigger than the PBE results and are 2.53, 2.18 and 1.47 eV for ZnS$_2$, CdS$_2$ and CdSe$_2$, respectively, which are in good agreement with the band gaps calculated by a previous GW study.[30] Therefore, in this work, in order to evaluate the electrical transport properties of the three dichalcogenides more accurately, we manually change their band gaps to their HSE06-calculated values, but keep the PBE-calculated band structure shapes. Afterwards, by solving the electron BTE under the CRTA, the electrical transport properties of the three dichalcogenides at different carrier concentrations and temperatures are evaluated with the BoltzTraP2 code[25], and the calculated Seebeck coefficients ($S$) and electrical conductivities with respect to relaxation time ($\sigma/\tau$) in ZnS$_2$, CdS$_2$ and CdSe$_2$ as a function of carrier concentration at different temperatures (300, 500 and 700 K) are shown in Fig. 7a and Fig. 7b, respectively.

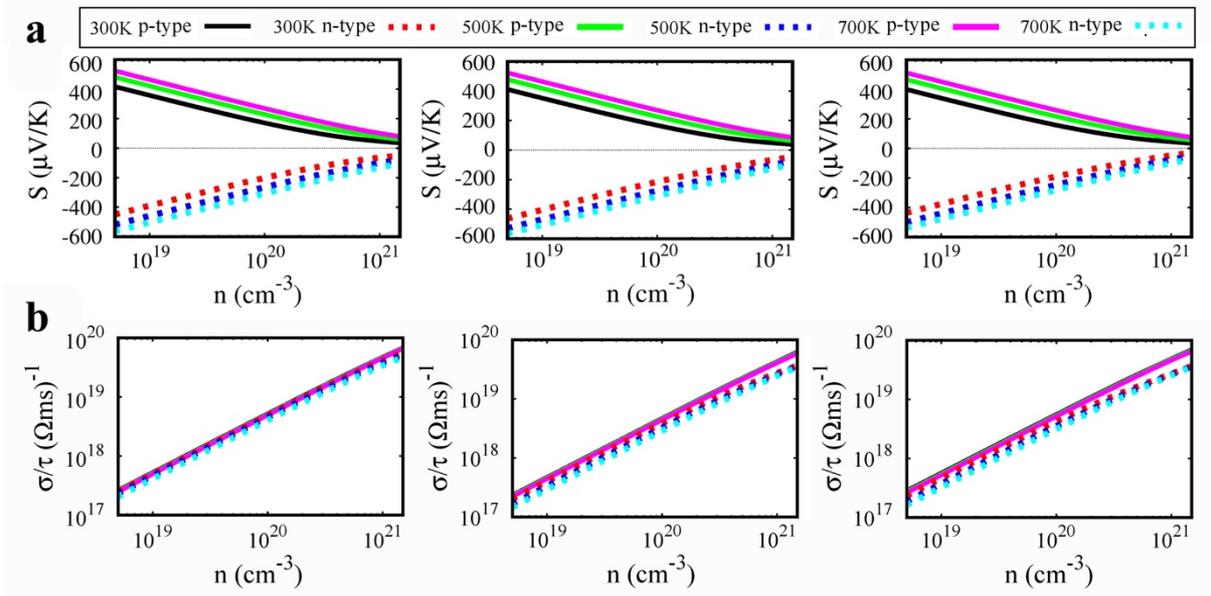

Figure 7. Calculated Seebeck coefficients $S$ (a) and electrical conductivities with respect to relaxation time $\sigma/\tau$ (b) in ZnS$_2$, CdS$_2$ and CdSe$_2$ (from left to right) as a function of the carrier concentration for p-type and n-type doping at 300, 500 and 700 K, respectively.

From Fig. 7a, we can find that the calculated absolute values of $S$ ($|S|$) at 300 K with $10^{19}$ cm$^{-3}$ carrier concentration are 359, 353 and 342 μV/K for p-type carriers and 390, 405 and 377 μV/K for n-type carriers in ZnS$_2$, CdS$_2$ and CdSe$_2$, respectively, which are larger than that of the well-studied PbTe (~100 μV/K for n-type carriers)[45]. This means that all the three dichalcogenides have large Seebeck coefficients. Furthermore, we also find that their calculated $|S|$ are all decreased with the increase of the carrier concentration ($n$), which can be explained by the Mott formula, as[26]

$$S = \frac{2k_B^2}{3e\hbar^2} m_d^* \left(\frac{\pi}{3n}\right)^{\frac{2}{3}} T \qquad (2)$$

where $m_d^*$ is the density-of-states (DOS) effective mass, as $m_d^* = \Sigma_{N_V} \sqrt[3]{\Pi_{\alpha\beta} m_{\alpha\beta}^*}$. From Eq. 2, we can see that the values of $m_d^*$ have a significant effect on $S$. As reported in the previous researches[46], in a semiconductor with the degeneracy and non-spherical Fermi surface, $m_d^*$ is proportional to the Fermi surface complexity factor ($N_V K^*$), as $m_d^* \propto (N_V K^*)^{2/3}$, where $N_V$ is the energy valley degeneracy and $K^*$ is the anisotropy parameter. For the three dichalcogenides (Fig. 6), since their valence band shapes of B1 at the VBM are very flat along the Γ-X direction but very steep along the Γ–M and Γ-R directions, their isoenergy Fermi surfaces of B1 will be very complex and their $K^*$ will be very large for p-type doping. Besides, since their symmetry degeneracies at P point in CBM are large (8), their $N_V$ will be very large for n-type doping. Therefore, their large $K^*$ and $N_V$ can contribute to large $m_d^*$. Furthermore, by calculating $m_d^*$ using the data in Table 1, we can get that the calculated $m_d^*$ of the three dichalcogenides are $1.75m_0$, $1.65m_0$ and $1.53m_0$ for p-type carriers, and $4.05m_0$, $4.32m_0$ and $3.44m_0$ for n-type carriers, which have same trend with the calculated $|S|$. This further indicates that the large $|S|$ in the three dichalcogenides may be mainly contributed by their large $m_d^*$, and the larger $m_d^*$ can result in the larger $|S|$.

From Fig. 7b, we can find that the calculated $\sigma/\tau$ in the three dichalcogenides are all increased with the increase of carrier concentration ($n$), which is caused by the relationship between the electrical conductivity ($\sigma$) and $n$, as[26]

$$\sigma = \frac{ne^2}{m_c^*}\tau \qquad (3)$$

where $m_c^*$ is the conductivity effective mass, as $m_c^* = \left[\frac{1}{3}\Sigma_{\alpha\beta}\frac{1}{m_{\alpha\beta}^*}\right]^{-1}$. From Eq. 3, we can know that the values of $\sigma/\tau$ are mainly decided by $m_c^*$ and $m_{\alpha\beta}^*$. From the above analyses of electronic band structures, we can gather that due to the small $m_{\alpha\beta}^*$ along different directions of B2 band at the VBM and B band at the CBM (in Table 1), all the three dichalcogenides have light p-type and n-type carriers, which can contribute to the good $\sigma/\tau$ for both p-type and n-type doping in the three dichalcogenides. Furthermore, by obtaining $m_c^*$ at different carrier concentrations and temperatures (as shown in Fig. S3), we find that the calculated $m_c^*$ of $CdS_2$ and $CdSe_2$ for n-type doping are increased more obviously with the increase of temperature and carrier concentration than $ZnS_2$, which can be used to explain why the calculated $\sigma/\tau$ of $ZnS_2$ for n-type doping are better than that of $CdS_2$ and $CdSe_2$ at high temperatures and high carrier concentrations (as shown in Fig. 7b).

To complete the evaluation of the electrical transport properties of the three dichalcogenides, we use the simple CRTA (by assuming $\tau = 10^{-14}$ s at different carrier concentrations and temperatures) to evaluate the carrier relaxation times ($\tau$). To justify the CRTA, we also tried a model similar to that used for ZnSe$_2$ and the result showed that the calculated $\tau$ was independent of the carrier concentration, which means the CRTA model is reasonable for the three dichalcogenides.[17] Combining the calculated large $S$ and good $\sigma/\tau$, the calculated thermopower factors ($PF = S^2\sigma$) in the three dichalcogenides as a function of carrier concentration at different temperatures are shown in Fig. 8. From it, we can find that the calculated $PF$ are first increased and then decreased with increasing carrier concentration, and the maximum $PF$ ($PF_{max}$) at 300 K can reach 1.56, 1.36 and 1.42 mW/(m·K$^2$) for p-type carriers, and 2.24, 1.97 and 1.59 mW/(m·K$^2$) for n-type carriers in ZnS$_2$, CdS$_2$ and CdSe$_2$, respectively, which are comparable with the well-known thermoelectric material SnSe [1.01 mW/(m·K$^2$)][37]. This means that all the three dichalcogenides have promising electrical transport properties. Furthermore, since ZnS$_2$ has the larger $\sigma/\tau$, the calculated $PF_{max}$ for n-type carriers at high temperatures in ZnS$_2$ are higher than those in CdS$_2$ and CdSe$_2$.

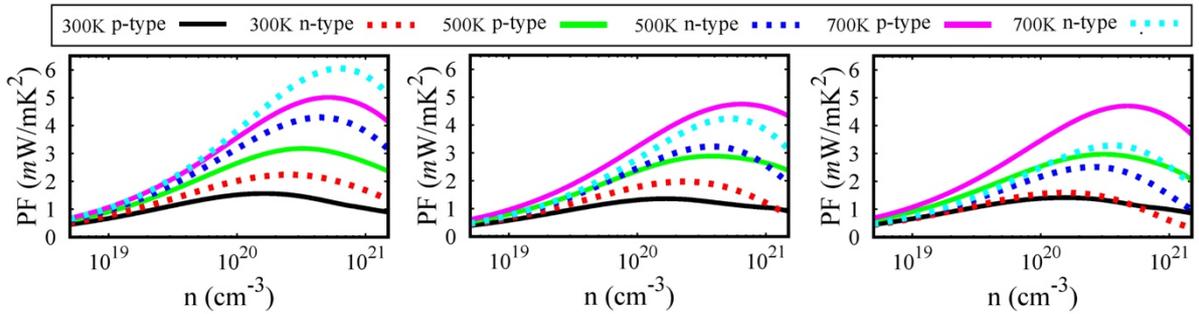

Figure 8. The thermopower factors ($PF$) in ZnS$_2$, CdS$_2$ and CdSe$_2$ (from left to right) as a function of the carrier concentration for p-type and n-type doping at 300, 500 and 700 K, respectively.

### F.  Figures of merits

Combining the calculated thermal conductivity ($\kappa_l$ and $\kappa_e$) and electrical transport properties ($PF$), the figures of merit ($ZT$) of the three dichalcogenides (ZnS$_2$, CdS$_2$ and CdSe$_2$) are evaluated as a function of the carrier concentration at 300, 500 and 700 K, respectively, as shown in Fig. 9. From it, we can see that the maximum $ZT$ ($ZT_{max}$) of the three dichalcogenides at 700 K are 1.98, 1.89 and 2.81 at 7.89×10$^{19}$, 7.73×10$^{19}$ and 4.70×10$^{19}$ cm$^{-3}$

for p-type carriers, and 2.42, 2.19 and 2.82 at $1.09\times10^{20}$, $1.40\times10^{20}$ and $7.19\times10^{19}$ cm$^{-3}$ for n-type carriers, all higher than 1.5. This indicates that the three dichalcogenides can be the excellent thermoelectric materials for both p-type and n-type doping, and they are worth for the future experimental investigations. Furthermore, due to the higher *PF* for n-type doping in ZnS$_2$ and the lower $\kappa_l$ in CdSe$_2$, the calculated $ZT_{max}$ for n-type doping in ZnS$_2$ and the calculated $ZT_{max}$ for both p-type and n-type doping in CdSe$_2$ are higher than those in CdS$_2$.

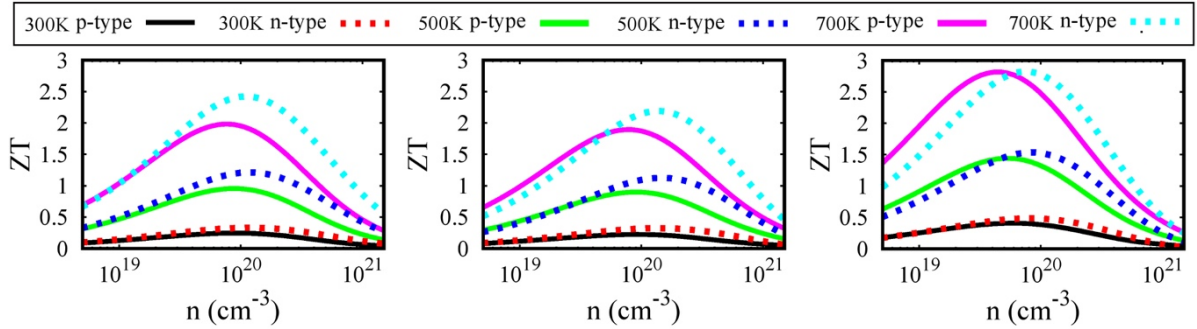

Figure 9. Calculated *ZT* values of ZnS$_2$, CdS$_2$ and CdSe$_2$ (from left to right) as a function of the carrier concentration for p-type and n-type doping at 300, 500 and 700 K, respectively.

## CONCLUSIONS

In this work, by solving the phonon and electron Boltzmann transport equations, we systematically study the thermal and electrical transport properties of three pyrite-type IIB-VIA$_2$ dichalcogenides (ZnS$_2$, CdS$_2$ and CdSe$_2$). The detailed analyses of phonon properties explain that there are loosely bound rattling-like metal atoms in all three, and the heavy atom masses, large atomic displacement parameters and long bond lengths between metal and nonmetal atoms can be beneficial to the rattling behaviors of the metal atoms and soft phonon modes. The evaluations and analyses of the Grüneisen parameters indicate that the three dichalcogenides exhibit very strong anharmonicity, traceable to the vibrations of all atoms perpendicular to the strongly bound nonmetallic dimers. The soft phonon modes and strong anharmonicities result in low lattice thermal conductivities in the three pyrite-type dichalcogenides. The calculations of electrical transport properties show that the three dichalcogenides have similar electronic band structures, and all of them can possess large DOS effective masses and small conductivity effective masses for both p-type and n-type carriers. The large DOS effective masses can contribute to their high Seebeck coefficients, while the small conductivity effective masses can be beneficial to their good electrical conductivities. Their high Seebeck coefficients and good electrical conductivities guarantee their promising electrical transport properties for both p-type and n-type carriers. Combining their low thermal conductivities and promising electrical transport properties, their maximum *ZT* values at 700 K can reach 1.98, 1.89 and 2.81 for p-type doping, and 2.42, 2.19 and 2.82 for n-type doping, respectively. Our work shows that the three pyrite-type dichalcogenides ($ZnS_2$, $CdS_2$ and $CdSe_2$) all exhibit excellent thermoelectric properties for both p-type and n-type doping. Combine with previous research about $ZnSe_2$, which has *ZT* values of 2.21 for p-type doping and 1.87 for n-type doping at high temperatures[17], we can infer that the pyrite-type $IIB-VIA_2$ dichalcogenides have promising thermoelectric properties. Furthermore, we further suggest that alloying of these compounds could be used to optimize their thermoelectric properties in the future.

## ACKNOWLEDGMENTS

This work is supported by National Natural Science Foundation of China, Grant Nos. 11774347, 12088101, 11991060, 12104035, and U1930402, and by China Postdoctoral Science Foundation, Grant No. 2021M690327, and by China Scholarship Council (CSC).